\documentclass[conference]{IEEEtran}
\pdfoutput=1
\makeatletter
\def\bstctlcite{\@ifnextchar[{\@bstctlcite}{\@bstctlcite[@auxout]}}
\def\@bstctlcite[#1]#2{\@bsphack
 \@for\@citeb:=#2\do{%
   \edef\@citeb{\expandafter\@firstofone\@citeb}%
   \if@filesw\immediate\write\csname #1\endcsname{\string\citation{\@citeb}}\fi}%
 \@esphack}
\makeatother

\IEEEoverridecommandlockouts %
\usepackage[cmex10]{amsmath} %
\usepackage{amssymb} %
\usepackage{bm}      %

\usepackage{url}
\usepackage{ifpdf}
\ifpdf
\usepackage[pdftex]{graphicx,color} %
\usepackage{float} %
\usepackage{hyperref} %
\else
\usepackage[dvips]{graphicx,color}
\usepackage{epsfig} %
\fi

\usepackage{algorithm} %
\usepackage{algpseudocode}

\usepackage{fixltx2e} %
\usepackage[caption=false,font=small]{subfig}

\def\Cc{{\mathcal C}}
\def\Dc{{\mathcal D}}
\def\Ec{{\mathcal E}}

\def\Jc{{\mathcal J}}

\def\Lc{{\mathcal L}}

\def\Sc{{\mathcal S}}

\def\Uc{{\mathcal U}}
\def\Vc{{\mathcal V}}
\def\Xc{{\mathcal X}}
\newcommand{\vect}[1]{\bm{#1}} %

\DeclareMathOperator*{\argmin}{arg\,min}
\def\registered{{\ooalign{\hfil\raise .00ex\hbox{\scriptsize R}\hfil\crcr\mathhexbox20D}}}

\newtheorem{property}{Property}

\newenvironment{definition}[1][Definition]{\begin{trivlist}
\item[\hskip \labelsep \textsc{#1}]}{\end{trivlist}}

\newcommand{\onefigure}[6]{
  \begin{figure}[t!]  \begin{center} \includegraphics[height=#2,width=#3,keepaspectratio=true]{#1} \vspace{-3mm} \caption[#4]{{\small
  #5}} \label{#6} \vspace{-4mm} \end{center} \end{figure} }

\usepackage[ireport]{KTHEEtitlepage}

\graphicspath{{figures/}}
\ifpdf
\DeclareGraphicsExtensions{.pdf,.jpeg,.png}
\else
\DeclareGraphicsExtensions{.eps}
\fi

\begin{document}

\ititle{Multi-path Routing Metrics for\\
       Reliable Wireless Mesh Routing Topologies}
\isubtitle{This work was supported in part by HSN (Heterogeneous Sensor
Networks), which receives support from Army Research Office (ARO)
Multidisciplinary Research Initiative (MURI) program (Award number
W911NF-06-1-0076) and in part by TRUST (Team for Research in
Ubiquitous Secure Technology), which receives support from the
National Science Foundation (NSF award number CCF-0424422) and the
following organizations: AFOSR (\#FA9550-06-1-0244), BT, Cisco, DoCoMo
USA Labs, EADS, ESCHER, HP, IBM, iCAST, Intel, Microsoft, ORNL,
Pirelli, Qualcomm, Sun, Symantec, TCS, Telecom Italia, and United
Technologies.  The work was also supported by the EU project FeedNetBack,
the Swedish Research Council, the Swedish Strategic Research Foundation,
and the Swedish Governmental Agency for Innovation Systems.}
\iauthor{Phoebus Chen, Karl H. Johansson, Paul Balister,
         B\'{e}la Bollob\'{a}s, and Shankar Sastry}
\idate{2011}
\irefnr{TRITA-EE 2011:033}

\iaddress{ACCESS Linnaeus Centre\\
  Automatic Control\\
  School of Electrical Engineering\\
  KTH Royal Institute of Technology\\
  SE-100 44 Stockholm, Sweden}
\makeititle

\bstctlcite{IEEEtranBSTcontrol} %

\title{Multi-path Routing Metrics for\\
       Reliable Wireless Mesh Routing Topologies}

\author{\IEEEauthorblockN{Phoebus Chen and Karl H. Johansson}
\IEEEauthorblockA{ACCESS Linnaeus Centre\\
KTH Royal Institute of Technology\\
Stockholm, Sweden\\
}
\and
\IEEEauthorblockN{Paul Balister and B\'{e}la Bollob\'{a}s}
\IEEEauthorblockA{Department of Mathematical Sciences\\
University of Memphis, USA\\
}
\and
\IEEEauthorblockN{Shankar Sastry}
\IEEEauthorblockA{Department of EECS\\
University of California, Berkeley, USA\\
}
}

\maketitle

\begin{abstract}
Several emerging classes of applications that run over wireless
networks have a need for mathematical models and tools to
systematically characterize the reliability of the network.  We
propose two metrics for measuring the reliability of wireless mesh
routing topologies, one for flooding and one for unicast routing.  The
Flooding Path Probability (FPP) metric measures the end-to-end packet
delivery probability when each node broadcasts a packet after hearing
from all its upstream neighbors.  The Unicast Retransmission Flow
(URF) metric measures the end-to-end packet delivery probability when
a relay node retransmits a unicast packet on its outgoing links until
it receives an acknowledgement or it tries all the links.  Both
metrics rely on specific packet forwarding models, rather than
heuristics, to derive explicit expressions of the end-to-end packet
delivery probability from individual link probabilities and the
underlying connectivity graph.

We also propose a distributed, greedy algorithm that uses the URF
metric to construct a reliable routing topology.  This algorithm
constructs a Directed Acyclic Graph (DAG) from a weighted, undirected
connectivity graph, where each link is weighted by its success
probability.  The algorithm uses a vector of decreasing reliability
thresholds to coordinate when nodes can join the routing topology.
Simulations demonstrate that, on average, this algorithm constructs a
more reliable topology than the usual minimum hop DAG.
\end{abstract}


\IEEEpeerreviewmaketitle

\section{Introduction}
\label{sec:introduction}

Despite the lossy nature of wireless channels, applications that need
reliable communications are migrating toward operation over wireless
networks.  Perhaps the best example of this is the recent push by the
industrial automation community to move part of the control and
sensing infrastructure of networked control systems (see
\cite{hespanha-procIEEE:2007} for a survey of the field) onto Wireless
Sensor Networks (WSNs) \cite{WINA:url,RFC5673:2009}.  This has
resulted in several efforts to create WSN communication standards
tailored to industrial automation (e.g., WirelessHART
\cite{chen-book:2010}, ISA-SP100 \cite{ISA-SP100:url}).

A key network performance metric for all these communication standards
is \emph{reliability}, the probability that a packet is successfully
delivered to its destination.  The standards use several mechanisms to
increase reliability via diversity, including retransmissions (time
diversity), transmitting on different frequencies (frequency
diversity), and multi-path routing (spatial / path diversity).  But
just providing mechanisms for higher reliability is not enough ---
methods to characterize the reliability of the network are also needed
for optimizing the network and for providing some form of performance
guarantee to the applications.  More specifically, we need a network
reliability metric in order to: 1) quickly evaluate and compare different
routing topologies to help develop wireless node deployment /
placement strategies; 2) serve as an abstraction / interface of the
wireless network to the systems built on these networks (e.g.,
networked control systems); and 3) aid in the construction of a
reliable routing topology.

This paper proposes two multi-path routing topology metrics, the
Flooding Path Probability (FPP) metric and the Unicast Retransmission
Flow (URF) metric, to characterize the reliability of wireless mesh
hop-by-hop routing topologies.  Both routing topology metrics are
derived from the directed acyclic graph (DAG) representing the routing
topology, the link probabilities (the link metric), and specific
packet forwarding models.  The URF and FPP metrics define different
ways of combining link metrics than the usual method of summing or
multiplying the link costs along single paths.

The merit of these routing topology metrics is that they clearly
relate the modeling assumptions and the DAG to the reliability of the
routing topology.  As such, they help answer questions such as: When
are interleaved paths with unicast hop-by-hop routing better than
disjoint paths with unicast routing?  Under what modeling assumptions
does routing on an interleaved multi-path topology provide better
reliability than routing along the best single path?  What network
routing topologies should use constrained flooding for good
reliability?  (These questions will be answered in
Sections~\ref{sec:fpp_discussion} and \ref{sec:urf_discussion}.)

Sections~\ref{sec:related_works} and \ref{sec:prob_description} provide
background on routing topology metrics and a more detailed problem
description, to better understand the contributions of this paper.

The contributions of this paper are two-fold: First, we define the FPP
and URF metrics and algorithms for computing them in
Sections~\ref{sec:fpp_metric} and \ref{sec:urf_metric}.  Second, we
propose a distributed, greedy algorithm called
\textsc{URF-Delayed\_Thresholds} (URF-DT) to generate a mesh routing
topology that locally optimizes the URF metric in
Section~\ref{sec:construct_route_topo}.  We demonstrate that the
URF-DT algorithm can build routing topologies with significantly
better reliability than the usual minimum hop DAG via simulations in
Section~\ref{sec:simulations}.

\section{Related Works}
\label{sec:related_works}

In single-path routing, the path metric is often defined as the sum of
the link metrics along the path.  Examples of link metrics include the
negative logarithm of the link probability (for path probability)
\cite{dubois-Ferriere:2010}, ETX (Expected Transmission Count), ETT
(Expected Transmission Time), and RTT (Round Trip Time)
\cite{campista:2008}.  Most single-path routing protocols find minimum
cost paths, where the cost is the path metric, using a shortest path
algorithm such as Dijkstra's algorithm or the distributed Bellman-Ford
algorithm \cite{CLR:1990}.

In multi-path routing, one wants metrics to compare collections of
paths or entire routing topologies with each other.  Simply defining
the multi-path metric to be the maximum or minimum single-path metric
of all the paths between the source and the sink is not adequate,
because such a multi-path metric will lose information about the
collection of paths.

Our FPP metric is a generalization of the reliability calculations
done in \cite{de:2003} for the M-MPR protocol and in \cite{ye-wn:2005}
for the GRAdient Broadcast protocol.  Unlike
\cite{de:2003,ye-wn:2005}, our algorithm for computing the FPP metric 
does not assume all paths have equal length.

Our URF metric is similar to the anypath route metric proposed by
dubois-Ferriere et al. \cite{dubois-Ferriere:2010}.  Anypath routing,
or opportunistic routing, allows a packet to be relayed by one of
several nodes which successfully receives a packet \cite{biswas:2004}.
The anypath route metric generalizes the single-path metric by
defining a ``links metric'' between a node and a set of candidate
relay nodes.  The specific ``links metric'' is defined by the
candidate relay selection policy and the underlying link metric (e.g.,
ETX, negative log link probability).  As explained later in
Section~\ref{sec:urf_discussion}, although the packet forwarding
models for the URF and FPP metrics are not for anypath routing, a
variation of the URF metric is almost equivalent to the ERS-best E2E
anypath route metric presented in \cite{dubois-Ferriere:2010}.

One of our earlier papers, \cite{chen-robocomm:2007}, modeled the
precursor to the WirelessHART protocol, TSMP \cite{pister:2008}.  We
developed a Markov chain model to obtain the probability of packet
delivery over time from a given mesh routing topology and TDMA
schedule.  The inverse problem, trying to jointly construct a mesh
routing topology and TDMA schedule to satisfy stringent reliability
and latency constraints, is more difficult.  The approach taken in
this paper is to separate the scheduling problem from the routing
problem, and focus on the latter.  The works
\cite{soldati:2010,zou:2010} find the optimal schedule and packet
forwarding policies for delay-constrained reliability when given a
routing topology.

Many algorithms for building multi-path routing topologies try to
minimize single-path metrics.  For instance, \cite{sobrinho:2002}
extends Dijkstra's algorithm to find multiple equal-cost minimum cost
paths while \cite{bhandari:1997} finds multiple edge-disjoint and
node-disjoint minimum cost paths.  RPL \cite{IETF-RPL:2010v19}, a
routing protocol currently being developed by the IETF ROLL working
group, constructs a DAG routing topology by building a minimum cost
routing tree (links from child nodes to "preferred parent" nodes) and
then adding redundant links which do not introduce routing
loops.\footnote{The primary design scenario considered by RPL uses
  single-path metrics.  Other extensions to consider multi-path
  metrics may be possible in the future.}  In contrast, our URF-DT
algorithm constructs a reliable routing topology by locally optimizing
the URF metric, a multi-path metric that can express the reliability
provided by hop-by-hop routing over interleaved paths.

Another difference between URF-DT and RPL is that URF-DT specifies a
mechanism to control the order which nodes connect to the routing
topology, while RPL does not.  The connection order affects the
structure of the routing topology.

Finally, the LCAR algorithm proposed in \cite{dubois-Ferriere:2010}
for building a routing topology cannot be used to optimize the URF
metric because the underlying link metric (negative log link
probability) for the URF metric does not satisfy the physical cost
criterion defined in \cite{dubois-Ferriere:2010}.

\section{Problem Description}
\label{sec:prob_description}

We focus on measuring the reliability of wireless mesh routing
topologies for WSNs, where the wireless nodes have low computational
capabilities, limited memory, and low-bandwidth links to neighbors.

Empirical studies \cite{pister:2008} have shown that multi-path
hop-by-hop routing is more reliable than single-path routing in
wireless networks, where reliability is measured by the source-to-sink
packet delivery ratio.  The main problem is to define multi-path
reliability metrics for flooding and for unicast routing that capture
this empirical observation.  The second problem is to design an
algorithm to build a routing topology that directly tries to optimize
the unicast multi-path metric.

The FPP and URF metrics only differ in their packet forwarding models,
which are discussed in Sections~\ref{sec:fpp_fwd_model} and
\ref{sec:urf_fwd_model}.  Both models do not retransmit packets on
failed links.  More accurately, a finite number of retransmissions on
the same link can be treated as one link transmission with a higher
success probability.\footnote{We can do this because our metrics only
measure reliability and are not measuring throughput or delay.}
Here, a failed link in the model describes a link outage that is
longer than the period of the retransmissions (a bursty link).

In fact, without long link outages and finite retransmissions, it is
hard to argue that multi-path hop-by-hop routing has better
reliability than single-path routing.  Under a network model where all
the links are mutually independent and independent of their past
state, all single paths have reliability 1 when we allow for an
infinite number of retransmissions.

Both the FPP and URF metrics assume that the links in the network
succeed and fail independently of each other.  While this is not
entirely true in a real network, it is more tractable than trying to
model how links are dependent on each other.  Both metrics also assume
that each node can estimate the probability that an incoming or
outgoing link fails through link estimation techniques at the link and
physical layers \cite{baccour:2009}.

\subsection{Notation and Terminology}
\label{sec:note_term}

We use the following notation and terminology to describe graphs.  Let
$G = (\Vc,\Ec,p)$ represent a weighted directed graph with the set of
vertices (nodes) $\Vc = \{1,\ldots,N\}$, the set of directed edges
(links) $\Ec \subseteq \{(i,j) \: : \: i,j \in \Vc\}$, and a
function assigning weights to edges $p : \Ec \mapsto [0,1]$.  The edge
weights are link success probabilities, and for more compact
notation we use $p_l$ or $p_{ij}$ to denote the probability of link $l
= (i,j)$.  The number of edges in $G$ is denoted $E$.  In a similar
fashion to $G$, let $\bar{G} = (\bar{\Vc},\bar{\Ec},p)$ represent a
weighted undirected graph (but now $\bar{\Ec}$ consists of undirected
edges).

The source node is denoted $a$ and the sink (destination) node is
denoted $b$.  A vertex cut of $a$ and $b$ on a connected graph is a
set of nodes $\Cc$ such that the subgraph induced by $\Vc \backslash
\Cc$ does not have a single connected component that contains both $a$
and $b$.  Note that this definition differs from the conventional
definition of a vertex cut because $a$ and $b$ can be elements in
$\Cc$.

The graph $G$ is a \emph{DODAG} (Destination-Oriented DAG) if all the
nodes in $G$ have at least
one outgoing edge except for the destination node $b$, which has no
outgoing edges.  We say that a node $i$ is \emph{upstream} of a node
$j$ (and node $j$ is \emph{downstream} of node $i$) if there exists a
directed path from node $i$ to $j$ in $G$.  Similarly, node $i$ is an
\emph{upstream neighbor} of node $j$ (and node $j$ is a
\emph{downstream neighbor} of node $i$) if $(i,j)$ is an edge in
$\Ec$.  The indegree of a node $i$, denoted as $\delta^-(i)$, is the
number of incoming links, and similarly the outdegree of a node $i$,
denoted as $\delta^+(i)$, is the number of outgoing links.  The
maximum indegree of a graph is $\Delta^- = \max_{i \in \Vc}
\delta^-(i)$ and the maximum outdegree of a graph is $\Delta^+ =
\max_{i \in \Vc} \delta^+(i)$.

Finally, define $2^\Xc$ to be the set of all subsets of the set $\Xc$.

\section{FPP Metric}
\label{sec:fpp_metric}

This section presents the FPP metric, which assumes that multiple
copies of a packet are flooded over the routing topology to try all
possible paths to the destination.

\subsection{FPP Packet Forwarding Model}
\label{sec:fpp_fwd_model}

In the FPP packet forwarding model, a node listens for a packet from
all its upstream neighbors and multicasts the packet once on all its
outgoing links once it receives a packet.  There are no retransmissions
on the outgoing links even if the node receives multiple copies of the
packet.  The primary difference between this forwarding model and
general flooding is that the multicast must respect the orientation of
the edges in the routing topology DAG.

\subsection{Defining and Computing the Metric}
\label{sec:fpp_def_comp_metric}

\begin{definition}
\label{def:fpp_metric}
\textbf{Flooding Path Probability Metric}\\
Let $G = (\Vc,\Ec,p)$ be a weighted DODAG, where each link $(i,j)$ in
the graph has a probability $p_{ij}$ of successfully delivering a
packet and all links independently succeed or fail.  The \emph{FPP}
metric $p_{a \rightarrow b} \in [0,1]$ for a source-destination pair
$(a,b)$ is the probability that a packet sent from node $a$ over the
routing topology $G$ reaches node $b$ under the FPP packet forwarding
model.
\IEEEQED
\end{definition}

Since the FPP packet forwarding model tries to send copies of the
packet down all directed paths in the network, $p_{a \rightarrow b}$
is the probability that a directed path of successful links exists in
$G$ between the source $a$ and the sink $b$.  This leads to a
straightforward formula to calculate the FPP metric.
\begin{equation}
\label{eq:fpp}
p_{a \rightarrow b} = \sum_{\Ec' \in 2_{a \rightarrow b}^\Ec} \left(
   \prod_{l \in \Ec'} p_l \prod_{\bar{l} \in \Ec \backslash \Ec'} (1-
   p_{\bar{l}}) \right) \qquad,
\end{equation}
where $2_{a \rightarrow b}^\Ec$ is the set of all subsets of $\Ec$ that
contain a path from $a$ to $b$.  Unfortunately, this formula is
computationally expensive because it takes $O(E 2^E)$ to compute.

Algorithm~\ref{alg:fpp} computes the FPP metric $p_{a \rightarrow b}$
using dynamic programming and is significantly faster.  The state used
by the dynamic programming algorithm is the joint probability
distribution of receiving a packet on vertex cuts $\Cc$ of the graph
separating $a$ and $b$ (See Figure~\ref{fig:fpp_vertex_cut_example}
for an example).  Recall that our definition of $\Cc$ allows $a$
and $b$ to be elements of $\Cc$, which is necessary for the first and
last steps of the algorithm.

\begin{algorithm}
\caption{\textsc{Fast\_FPP}}
\label{alg:fpp}
\begin{algorithmic}[5]
\State \textbf{Input}: $G=(\Vc,\Ec,p), a$ \Comment{$G$ is a connected DAG.}
\State \textbf{Output}: $\{p_{a \rightarrow v}, \forall v \in \Vc\}$
\State $\Cc := \{a\}$ \Comment{$\Cc$ is the vertex cut.}
\State $\Vc' := \Vc \backslash a$ \Comment{$\Vc'$ is the set of remaining vertices.}
\State $\Ec' := \Ec$ \Comment{$\Ec'$ is the set of remaining edges.}
\State $u := a$ \Comment{$u$ is the node targeted for removal from $\Cc$.}
\State $p_\Cc(\{a\}) := 1$; $p_\Cc(\emptyset) := 0$
       \Comment{pmf for vertex cut $\Cc$.}
\While{$\Vc' \ne \emptyset$}
  \State \hspace{-1em} \textbf{[Find node $u$ to remove from vertex cut]}
  \If{$u \not\in \Cc$} \label{line:add_node_policy}
    \State Let $\Jc = \{j \,:\, \forall (i,j) \in \Ec', i \in
           \Cc\}$ 
    \State $u := \argmin_{i \in \Cc} \big| \{ (i,j) \in \Ec' \,:\, j
           \in \Jc\} \big|$
  \EndIf
  \State \hspace{-1em} \textbf{[Add node $v$ to vertex cut]}
  \State Select any node $v \in \{j \in \Vc : (u,j) \in \Ec'\}$ 
         \label{line:add_node_policy2}
  \State $p_\Cc' := \mathtt{NIL}$ 
    \Comment{Probabilities for next vertex cut.}
  \ForAll{subsets $\Cc'$ of $\Cc$}
    \State Let $\Lc = \{(i,v) \in \Ec' \,:\, i \in \Cc'\}$
    \State $p_\Cc'(\Cc' \cup \{v\}) := p_\Cc(\Cc') \cdot 
           \left(1 - \prod_{l \in \Lc} (1- p_l) \right)$ \label{line:v_pkt}
    \State $p_\Cc'(\Cc') := p_\Cc(\Cc') \cdot \prod_{l \in \Lc} (1- p_l)$
           \label{line:v_no_pkt}
  \EndFor
  \State $\Ec' := \Ec' \backslash \{(i,v) \in \Ec' \,:\, i \in
          \Cc\}$
  \State $\Vc' := \Vc' \backslash v$
  \State $\Cc := \Cc \cup \{v\}$
  \State \hspace{-1em} \textbf{[Compute path probability]}
  \State Let $2_v^\Cc = \{\Cc' \in 2^\Cc \,:\, v \in \Cc'\}$
  \State $p_{a \rightarrow v} := \sum_{\Cc'_v \in 2_v^\Cc} p_\Cc'(\Cc'_v)$
         \label{line:sum_prob}
  \State \hspace{-1em} \textbf{[Remove nodes $\Dc$ from vertex cut]}
 \State Let $\Dc = \{i \in \Cc \,:\, \forall j, \, (i,j) \not\in \Ec'\}$ 
  \State $\Cc := \Cc \backslash \Dc$ 
  \State $p_\Cc := \mathtt{NIL}$
  \ForAll{subsets $\Cc'$ of $\Cc$}
    \State $p_\Cc(\Cc') := \sum_{\Dc' \in 2^\Dc} p_\Cc'(\Cc' \cup
    \Dc')$ \label{line:sum_prob2}
  \EndFor
\EndWhile
\State \textbf{Return}: $\{p_{a \rightarrow v}, \forall v \in \Vc\}$
\end{algorithmic}
\end{algorithm}

\onefigure{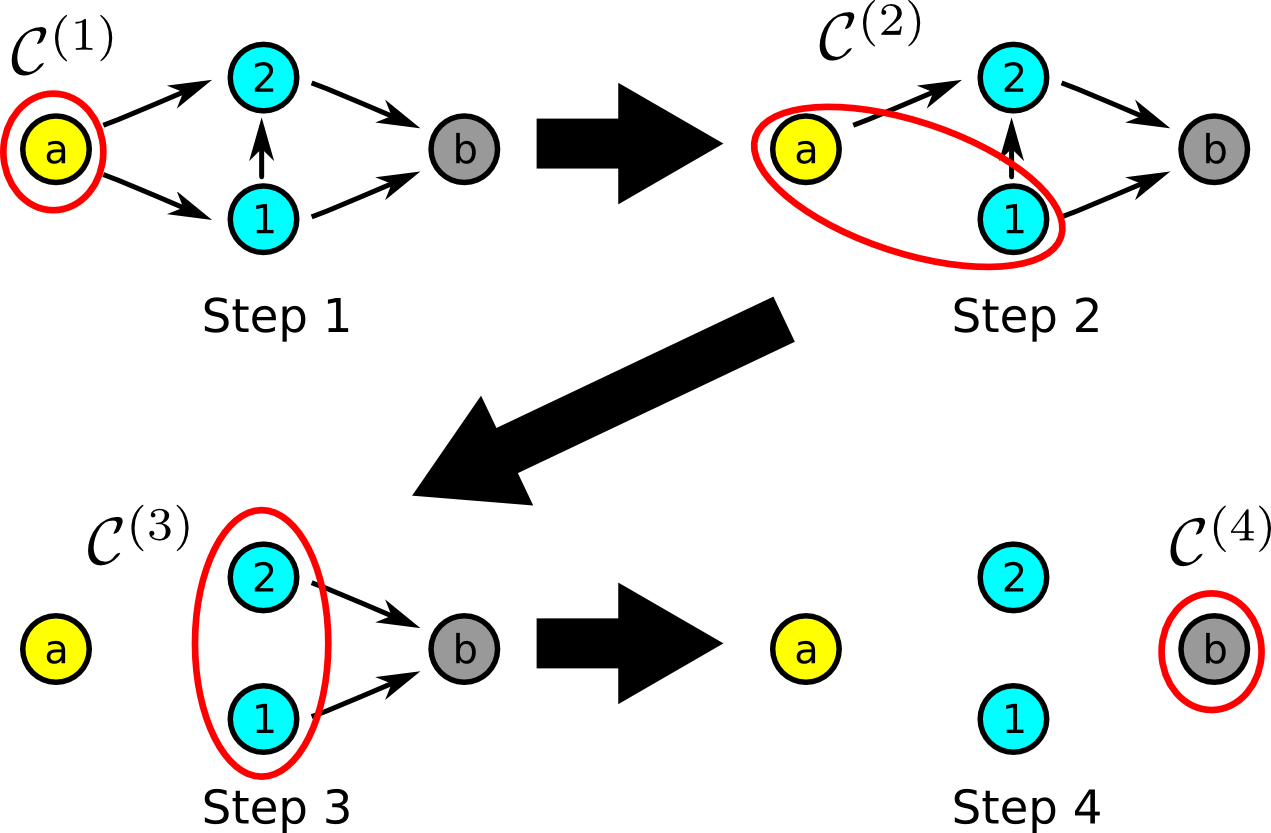}{!}{230pt}
	  {Example of vertex cuts for path probability}
	  {An example of a sequence of vertex cuts that can be used by
	   Algorithm~\ref{alg:fpp}.  The vertex cut after
	   adding and removing nodes from each iteration of the outer
	   loop is circled in red.}
	  {fig:fpp_vertex_cut_example}

\onefigure{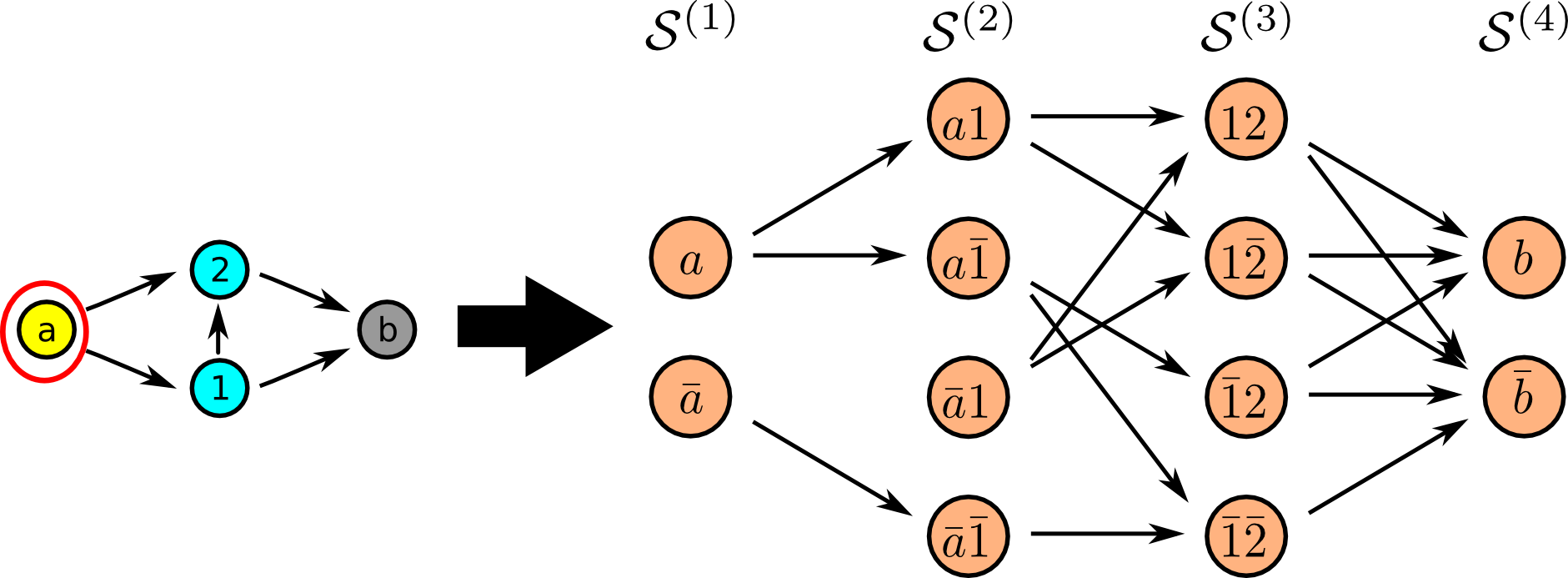}{!}{230pt}
	  {Example of vertex cut graph conversion for path
	   probability}
	  {Running Algorithm~\ref{alg:fpp} on the network graph
	   shown on the left when selecting vertex cuts in the order
	   depicted in Figure~\ref{fig:fpp_vertex_cut_example}
	   is equivalent to creating the vertex cut DAG shown on the
	   right and finding the probability that state $a$ will
	   transition to state $b$.}
	  {fig:fpp_graph_convert_example}

Conceptually, the algorithm is converting the DAG representing the
network to a \emph{vertex cut DAG}, where each vertex cut at step $k$,
$\Cc^{(k)}$, is represented by the set of nodes $\Sc^{(k)} =
2^{\Cc^{(k)}}$.  Each node in $\Sc^{(k)}$ represents the event that a
particular subset of the vertex cut received a copy of the packet.
The algorithm computes a probability for each node in $\Sc^{(k)}$, and
the collection of probabilities of all the nodes in $\Sc^{(k)}$
represent the joint probability distribution that nodes in the vertex
cut $\Cc^{(k)}$ can receive a copy of the packet.  A link in the
vertex cut DAG represents a valid (nonzero probability) transition
from a subset of nodes that have received a copy of the packet in
$\Cc^{(k-1)}$ to a subset of nodes that have received a copy of the
packet in $\Cc^{(k)}$.  Figure~\ref{fig:fpp_graph_convert_example}
shows an example of this graph conversion using the selection of
vertex cuts depicted in Figure~\ref{fig:fpp_vertex_cut_example}.

Algorithm~\ref{alg:fpp} tries to keep the vertex cut small by
using the greedy criteria in lines 
\ref{line:add_node_policy}--\ref{line:add_node_policy2}
to adds nodes to the vertex cut.  A node
can only be added to the vertex cut if all its incoming links
originate from the vertex cut.  When a node is added to the vertex
cut, its incoming links are removed.  A node is removed from the
vertex cut if all its outgoing links have been removed.

Computing the path probability $p_{a \rightarrow b}$ reduces to
computing the joint probability distribution that a packet is received
by a subset of the vertex cut in each step of the algorithm.  The
joint probability distribution over the vertex cut $\Cc^{(k)}$ is
represented by the function $p_\Cc^{(k)}:\Sc^{(k)} \mapsto [0,1]$.
Step $k$ of the algorithm computes $p_\Cc^{(k)}$ from $p_\Cc^{(k-1)}$
on lines \ref{line:v_pkt}, \ref{line:v_no_pkt}, and
\ref{line:sum_prob2} in Algorithm~\ref{alg:fpp}.  Notice that
the nodes in each $\Sc^{(k)}$ represent disjoint events, which is why
we can combine probabilities in lines \ref{line:sum_prob} and
\ref{line:sum_prob2} using summation.

\subsection{Computational Complexity}
\label{sec:fpp_comp_complex}

The running time of Algorithm~\ref{alg:fpp} is $O\big(N (\hat{C}
\Delta^+ + 2^{\hat{C}} \Delta^- )\big)$, where $\hat{C}$ is the size
of the largest vertex cut used in the algorithm.  This is typically
much smaller than the time to compute the FPP metric from
\eqref{eq:fpp}, especially if we restrict flooding to a subgraph of the
routing topology with a small vertex cut.  The analysis to get the
running time of Algorithm~\ref{alg:fpp} can be found in Section 2.2.2
of the dissertation \cite{chen:2009}.

The main drawback with the FPP metric is that it cannot be computed
in-network with a single round of local communication (i.e., between
1-hop neighbors).  Algorithm~\ref{alg:fpp} requires knowledge of the
outgoing link probabilities of a vertex cut of the network, but the
nodes in a vertex cut may not be in communication range of each other.
Nonetheless, if a gateway node can gather all the link probabilities
from the network, it can give an estimate of the end-to-end packet
delivery probability (the FPP metric) to systems built on this
network.

\subsection{Discussion}
\label{sec:fpp_discussion}

Figure~\ref{fig:wid3_len6_fpp_p0d7} shows the probability of nodes in
a mesh network receiving a packet flooded from the source.  This
simple topology shows that a network does not need to have large
vertex cuts to have good reliability in a network with poor links.  In
regions of poor connectivity, flooding constrained to a directed
acyclic subgraph with a small vertex cut can significantly boost
reliability.

\onefigure{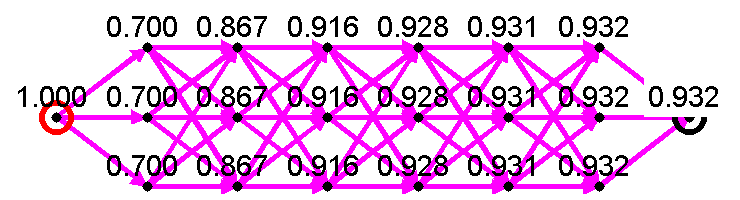}{!}{230pt}
          {FPP Metric Width-3 Length-6 Path}
          {FPP metric $p_{a \rightarrow v}$ for all nodes $v$, where
           all links have probability $0.7$.  The source node $a$ is
           circled in red.}
          {fig:wid3_len6_fpp_p0d7}

Oftentimes, it is not possible to estimate the probability of the
links accurately in a network.  Fortunately, since the FPP metric is
monotonically increasing with respect to all the link probabilities,
the range of the FPP metric can be computed given the range of each
link probability.  The upper (lower) bound on $p_{a \rightarrow b}$
can be computed by replacing every link probability $p_l$ with its
respective upper (lower) bound $\overline{p}_l$ ($\underline{p}_l$)
and running Algorithm~\ref{alg:fpp}.  For instance, the FPP metrics in
Figure~\ref{fig:wid3_len6_fpp_p0d7} can be interpreted as a lower
bound on the reliability between the source and each node if all links
have probability greater than $0.7$.

\section{URF Metric}
\label{sec:urf_metric}

This section presents the URF metric, which assumes that a single copy
of the packet is routed hop-by-hop over the routing topology.  Packets
are forwarded without prior knowledge of which downstream links have
failed.

\subsection{URF Packet Forwarding Model}
\label{sec:urf_fwd_model} 

Under the URF packet forwarding model, a node that receives a packet
will select a link uniformly at random from all its outgoing links for
transmission.  If the transmission fails, the node will select another
link for transmission uniformly at random from all its outgoing links
that \emph{have not been selected for transmission before}.  This
repeats until either a transmission on a link succeeds or the node has
attempted to transmit on all its outgoing links and failed each time.
In the latter case, the packet is dropped from the network.

\subsection{Defining and Computing the Metric}
\label{sec:urf_def_comp_metric}

\begin{definition}
\label{def:urf_metric}
\textbf{Unicast Retransmission Flow Metric}\\
Let $G = (\Vc,\Ec,p)$ be a weighted DODAG, where each link $(i,j)$ in
the graph has a probability $p_{ij}$ of successfully delivering a
packet and all links independently succeed or fail.  The \emph{URF}
metric $\varrho_{a \rightarrow b} \in [0,1]$ for a source-destination
pair $(a,b)$ is the probability that a packet sent from node $a$ over
the routing topology $G$ reaches node $b$ under the URF packet
forwarding model.
\IEEEQED
\end{definition}

The URF metric $\varrho_{a \rightarrow b}$ can be computed using
\begin{equation}
  \label{eq:urf_metric}
  \begin{aligned}
    \varrho_{a \rightarrow a} & = 1\\
    \varrho_{a \rightarrow v} & = \sum_{u \in \Uc_v}
       \varrho_{a \rightarrow u} \varpi_{uv} \qquad,
  \end{aligned}
\end{equation}
where $\Uc_v$ are all the upstream neighbors of node $v$ and
$\varpi_{uv} \in [0,1]$ is the Unicast Retransmission Flow
  weight (URF weight) of link $(u,v)$.  The URF weight for link $l =
(u,v)$ is the probability that a packet at $u$ will traverse the link
to $v$, and is given by
\begin{equation}
  \begin{aligned}
    \label{eq:urf_weight}
      \varpi_{uv} = \sum_{\Ec' \in 2^{\Ec_u \backslash l}} 
        \frac{p_{uv}}{|\Ec'|+1} \left( \prod_{e \in \Ec'} p_e \right)
        \left( \prod_{\bar{e} \in \Ec_u \backslash (\Ec' \cup l)}
	       1 - p_{\bar{e}} \right) %
  \end{aligned}
\end{equation}
where $\Ec_u = \{(u,v) \in \Ec : v \in \Vc\}$ is the set of node $u$'s
outgoing links.

Next, we sketch how \eqref{eq:urf_metric} and \eqref{eq:urf_weight}
can be derived from the URF packet forwarding model.  Recall that only
one copy of the packet is sent through the network and the routing
topology is a DAG, so the event that the packet traverses link
$(u_1,v)$ is disjoint from the event that the packet traverses
$(u_2,v)$.  The probability that a packet sent from $a$ traverses link
$(u,v)$ is simply $\varrho_{a \rightarrow u}\varpi_{uv}$, where
$\varrho_{a \rightarrow u}$ is the probability that a packet sent from
node $a$ visits node $u$ (therefore, $\varrho_{a \rightarrow a} = 1$).
Thus, the probability that the packet visits node $v$ is the sum of
the probabilities of the events where the packet traverses an incoming
edge of node $v$, as stated in \eqref{eq:urf_metric}.

Now, it remains to show that $\varpi_{uv}$ as defined by
\eqref{eq:urf_weight} is the probability that a packet at $u$ will
traverse the link $(u,v)$.  Recall that a packet at $u$ will traverse
$(u,v)$ if all the previous links selected by $u$ for transmission
fail and link $(u,v)$ is successful.  Alternately, this event
can be described as the union of several disjoint events arising from
two independent processes:
\begin{itemize}
\item each of $u$'s outgoing links is either up or down (with its
  respective probability), and
\item $u$ selects a link transmission order uniformly at random from
  all possible permutations of its outgoing links.
\end{itemize}
Each disjoint event is the intersection of: a particular realization
of the success and failure of $u$'s outgoing links where $(u,v)$ is
successful (corresponding to $p_{uv} \prod p_e \prod (1-p_{\bar{e}})$
in \eqref{eq:urf_weight}); and a permutation of the outgoing links
where $(u,v)$ is ordered before all the other successful links
(corresponding to $1/(|\Ec'|+1)$ in \eqref{eq:urf_weight}).  Summing
the probabilities of these disjoint events yields
\eqref{eq:urf_weight}.  For a rigorous derivation of the URF weights
from the packet forwarding model, please see Section 2.3.3 of the
dissertation \cite{chen:2009}.

\subsection{Computational Complexity}
\label{sec:urf_comp_complex}

The slowest step in computing the URF metric between all nodes and the
sink is computing \eqref{eq:urf_weight}, which has complexity
$O(\Delta^+\cdot2^{\Delta^+})$.  Using some algebra (See the
Appendix), \eqref{eq:urf_weight} simplifies to
\begin{equation}
    \label{eq:urf_weight2}
	   \varpi_{uv} = p_{uv} \int_0^1 \prod_{e \in \Ec_u \backslash (u,v)} 
           \left( 1 - p_e x \right)dx \qquad ,
\end{equation}
which can be evaluated efficiently in $O((\Delta^+)^2)$. This results
from the $O((\Delta^+)^2)$ operations to expand the polynomial and
$O(\Delta^+)$ operations to evaluate the integral.  Since there are
$O(\Delta^+)$ link weights per node and $N$ nodes in the graph, the
complexity to compute the URF metric sequentially on all nodes in the
graph is $O(N(\Delta^+)^3)$. (There are also $O(E)$ operations in
\eqref{eq:urf_metric}, but $E < 2N\Delta^+$.)  If we allow the link
weights to be computed in parallel on the nodes, then the complexity
becomes $O((\Delta^+)^3 + E)$.

Unlike the FPP metric, The URF metric can be computed in-network
with local message exchanges between nodes.  First, each node would
locally compute the URF link weights $\varpi_{uv}$ from link
probability estimates on its outgoing links.  Then, since the URF
metric $\varrho_{a \rightarrow v}$ is a linear function of the URF
weights, we can rewrite \eqref{eq:urf_metric} as 
\begin{equation}
  \label{eq:urf_metric_ref}
  \begin{aligned}
    \varrho_{b \rightarrow b} & = 1\\
    \varrho_{u \rightarrow b} & = \sum_{v \in \Vc_u}
       \varpi_{uv} \varrho_{v \rightarrow b} \qquad,
  \end{aligned}
\end{equation}
where $\Vc_u$ are all the downstream neighbors of node $u$.  This
means that each node $u$ only needs the URF metric of its downstream
neighbors to compute its URF metric to the sink, so the calculations
propagate outwards from the sink with only one message exchange on
each link in the DAG.

\subsection{Discussion}
\label{sec:urf_discussion}

The URF forwarding model can be implemented in both CSMA and TDMA
networks.  In the latter it describes a randomized schedule that is
agnostic to the quality of the links and routes in the network, such
that the scheduling problem is less coupled to the routing problem.
Loosely speaking, such a randomized packet forwarding policy is also
good for load balancing and exploiting the path diversity of mesh
networks.

The definition of the URF link weights is tightly tied to the URF
packet forwarding model.  One alternate packet forwarding model would
be for a node to always attempt transmission on outgoing links $(u,v)$
in decreasing order of downstream neighbor URF metrics $\varrho_{v
  \rightarrow b}$.  As before, the node tries each link once and drops
the packet when all links fail.\footnote{An opportunistic packet
forwarding model that would result in the same metric would
broadcast the packet once and select the most reliable relay to continue
forwarding the packet.}  This model leads to the following
Remaining-Reliability-ordered URF metric (RRURF), $\varrho_{a
\rightarrow b}'$, also calculated like $\varrho_{a \rightarrow b}$
from \eqref{eq:urf_metric_ref} except $\varpi_{uv}$ is replaced by
\begin{equation}
  \begin{aligned}
    \label{eq:rrurf_weight}
      \varpi_{l_i}' = \prod_{k=1}^{i-1} (1-p_{l_k})p_{l_i} \quad ,
  \end{aligned}
\end{equation}
where the outgoing links of node $u$ have been sorted into the list
$(l_1,\ldots,l_{\delta^+(u)})$ from highest to lowest downstream
neighbor URF metrics.\footnote{The RRURF metric would be equivalent to
the ERS-best E2E anypath routing metric of \cite{dubois-Ferriere:2010}
if every $D_x$ in the remaining path cost $R_{iJ}^{\mathrm{best}}$
(Equation 5 in \cite{dubois-Ferriere:2010}) were replaced by $\mathrm{exp}(-D_x)$.}

Notice that with unicast, a packet can reach a node where all its
outgoing links fail, i.e., the packet is ``trapped at a node.''  Thus,
topologies where a node is likely to receive a packet but has outgoing
links with very low success probabilities tend to perform poorly.
Flooding is not affected by this phenomenon of ``trapped packets''
because other copies of the packet can still propagate down other
paths.  In fact, given the same routing topology $G$, the URF metric
$\varrho_{a \rightarrow v}$ is always less than the FPP metric $p_{a
  \rightarrow v}$ for all nodes $v$ in the network.  The URF and FPP
metrics allow us to compare how much reliability is lost when
unicasting packets.  A comparison of
Figure~\ref{fig:wid3_len6_urf_p0d7} with
Figure~\ref{fig:wid3_len6_fpp_p0d7} reveals that this drop in
reliability can be significant in deep networks with low probability
links.  Nonetheless, unicast routing over a mesh still provides much
better reliability than routing down a single path or a small number
of disjoint paths with the same number of hops and the same link
probabilities, if the links are independent and bursty.

\onefigure{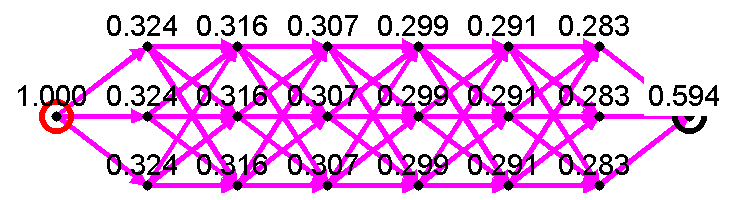}{!}{230pt}
          {URF Metric Width-3 Length-6 Path}
          {URF metric $\varrho_{a \rightarrow v}$ for all nodes $v$,
           where all links have probability $0.7$.  The source node
           $a$ is circled in red.} 
          {fig:wid3_len6_urf_p0d7}

Below are several properties of the URF metric that will be exploited
in Section~\ref{sec:construct_route_topo} to build a good mesh routing
topology.
 
\begin{property}[Trapped Packets]
\label{prop:urf_add_link_lower}
Adding an outgoing link to a node can lower its URF metric.
Similarly, increasing the probability of an outgoing link can also
lower a node's URF metric. \IEEEQED
\end{property}

Property~\ref{prop:urf_add_link_lower} can be seen on the example
shown in Figure~\ref{fig:urf_3_node_example1}. Here, link $(2,1)$
lowers the reliability of node $2$ to $b$.  Generally, nodes want to
route to other nodes that have better reliability to the sink, but
Figure~\ref{fig:urf_3_node_example2} shows an example where routing to
a node with worse reliability can increase your reliability.

{\captionsetup[subfloat]{position=top}
\begin{figure}
\subfloat[Illustration of Property~\ref{prop:urf_add_link_lower}.]{
\begin{minipage}[t]{110pt}
  \label{fig:urf_3_node_example1}
  \centering
  \includegraphics[width=90pt]{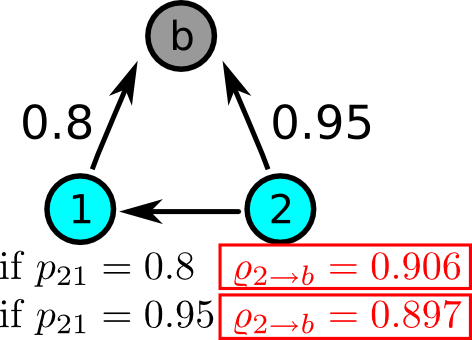}
\end{minipage} } \hfill
\subfloat[Illustration of Property~\ref{prop:urf_add_link_raise}.]{
\begin{minipage}[t]{110pt}
  \label{fig:urf_3_node_example2}
  \centering
  \includegraphics[width=90pt]{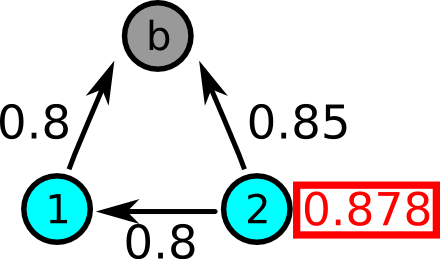}
\end{minipage} }
\caption{ Links are labeled with probabilities, and nodes are labeled
  with URF metrics $\varrho_{v \rightarrow b}$ (boxed).
  \subref{fig:urf_3_node_example1} Increasing $p_{12}$ lowers node 2's reliability.
  \subref{fig:urf_3_node_example2} Node 1 has a lower probability link
  to the sink than node 2, but link $(2,1)$ boosts the reliability of
  node 2.}
\label{fig:urf_3_node_examples}
\end{figure}
}

\begin{property}
\label{prop:urf_add_link_raise}
A node $u$ may add an outgoing link to node $v$, where $\varrho_{v
\rightarrow b} < \varrho_{u \rightarrow b}$, to increase $u$'s
URF metric. \IEEEQED
\end{property}

Property~\ref{prop:urf_add_link_raise} means that adding links
between nodes with poor reliability to the sink can boost their
reliability, as shown in Figure~\ref{fig:urf_4clique_1sink_example}.

\onefigure{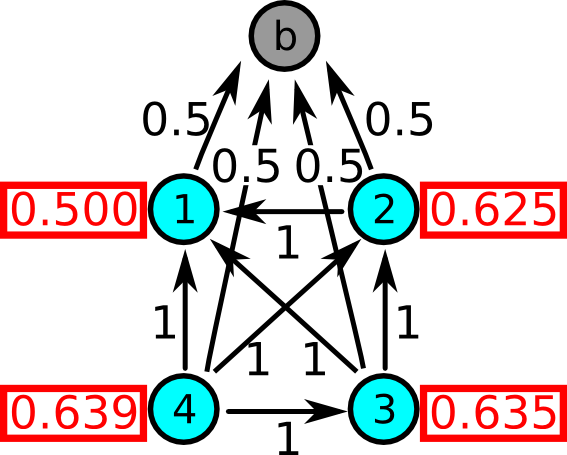}
          {70pt}{!}
          {Cross links increase reliability of nodes}
          {Nodes can significantly increase their reliabilities using
           cross links.  Links are labeled with probabilities, and
           nodes are labeled with URF metrics $\varrho_{v \rightarrow
           b}$ (boxed).  Without the cross links (the links with
           probability 1 in the diagram), the nodes would all have
           URF metric $0.5$.}
          {fig:urf_4clique_1sink_example}

\begin{property}
\label{prop:increase_urf_downstream}
Increasing the URF metric of a downstream neighbor of node $u$ 
always increases $u$'s URF metric. \IEEEQED
\end{property}

Property~\ref{prop:increase_urf_downstream} is because $\varrho_{u
  \rightarrow b}$, defined by \eqref{eq:urf_metric}, is monotonically
increasing in $\varrho_{v \rightarrow b}$ for all $v$ that are
downstream neighbors of $u$.

\begin{property}
\label{prop:urf_vs_urf_downstream}
A node may have a greater URF metric than some of its downstream
neighbors (from Property~\ref{prop:urf_add_link_raise}), but not a
greater URF metric than all of its downstream neighbors. \IEEEQED
\end{property}

Property~\ref{prop:urf_vs_urf_downstream} comes from
\[
\varrho_{u \rightarrow b} = \sum_{v \in \Vc_u} \varpi_{uv} \varrho_{v \rightarrow b}
\le (\max_{v \in \Vc_u} \varrho_{v \rightarrow b}) \sum_{v \in \Vc_u}
  \varpi_{uv} \le \max_{v \in \Vc_u} \varrho_{v \rightarrow b} .
\]
Not surprisingly, Properties~\ref{prop:increase_urf_downstream} and
\ref{prop:urf_vs_urf_downstream} highlight the importance of ensuring
that nodes near the sink have a very high URF metric $\varrho_{v
 \rightarrow b}$ when deploying networks and building routing
topologies.

If there is uncertainty estimating the link probabilities, bounding
the URF metric is not as simple as bounding the FPP metric because the
URF metric is not monotonically increasing in the link probabilities,
as noted in Property~\ref{prop:urf_add_link_lower}.  However, the URF
metric $\varrho_{u \rightarrow b}$ is monotonically increasing with
the link flow weights $\varpi_{uv}$ so bounds on the flow weights can
be used to compute bounds on the URF metric by simple substitution.
Similarly, each flow weight varies monotonically with each link
probability, so it can also be bounded by simple substitution.  For
instance, to compute the upper bound of $\varpi_{uv}$, you would
substitute the upper bound $\overline{p_{uv}}$ for $p_{uv}$ and the
lower bounds $\underline{p_e}$ for all the other links in
\eqref{eq:urf_weight2}.  Note that the upper bounds for all the flow
weights on the outgoing links from a node may sum to a value greater
than 1, which would lead to poor bounds on the URF metric.

\section{Constructing a Reliable Routing Topology}
\label{sec:construct_route_topo}

The \textsc{URF-Delayed\_Thresholds} (URF-DT) algorithm presented
below uses the URF metric to help construct a reliable, loop-free
routing topology from an ad-hoc deployment of wireless nodes.  The
algorithm assumes that each node can estimate the packet delivery
probability of its links.  Only symmetric links, links where the
probability to send and receive a packet are the same, are used by the
algorithm.  The algorithm either removes or assigns an orientation to
each undirected link in the underlying network connectivity graph to
indicate the paths a packet can follow from its source to its
destination.  The resulting directed graph is the routing topology.

To ensure that the routing topology is loop-free, the URF-DT algorithm
assigns an ordering to the nodes and only allows directed edges from
larger nodes to smaller nodes.  The algorithm assigns a \emph{mesh hop
count} to each node to place them in an ordering, analogous to the
use of rank in RPL \cite{IETF-RPL:2010v19}.

The URF-DT algorithm is distributed on the nodes in the network and
constructs the routing topology (a DODAG) outward from the
destination.  Each node uses the URF metric to decide how to join the
network --- who it should select as its downstream neighbors such that
packets from the node are likely to reach the sink.  A node has an
incentive to join the routing topology after its neighbors have
joined, so they can serve as its downstream neighbors and provide more
paths to the sink.  To break the stalemate where each node is waiting
for another node to join, URF-DT forces a node to join the routing
topology if its reliability to the sink after joining would cross a
threshold.  This threshold drops over time to allow all nodes to
eventually join the network.

\subsection{URF Delayed Thresholds Algorithm}
\label{sec:delay_thresh}

The URF-DT algorithm given in Algorithm~\ref{alg:delay_thresh}
operates in rounds, where each round lasts a fixed interval of time.
The algorithm requires all the nodes share a global time (e.g., by a
broadcast time synchronization algorithm) so they can keep track
of the current round $k$.

\begin{algorithm}
\caption{\textsc{URF-Delayed\_Thresholds}}
\label{alg:delay_thresh}
\begin{algorithmic}
\State \textbf{Input}: connectivity graph 
                       $\bar{G}=(\bar{\Vc},\bar{\Ec},p), b, \vect{\tau}, K$
\State \textbf{Output}: \parbox[t]{6.5cm}
                        {routing topology $G=(\Vc,\Ec,p)$,\\
                         mesh hop counts $\vect{\hbar}$}
\State $\Vc := \emptyset, \Ec := \emptyset$
\State $\forall i, \hbar_i := \mathtt{NIL}$
       \Comment{$\mathtt{NIL}$ means not yet assigned.} 
\State $\hbar_b := 0$
\For{$k := 1$ to $K$}
    \State \hspace{-1em}
    \textbf{\parbox[t]{8cm}{[Run this code simultaneously on all nodes $u \not\in \Vc$]}}
    \State Let $\hbar_{\Vc_u}^{\mathrm{min}} = \min_{v \in \Vc_u}\hbar_{v} \; , \; 
                \hbar_{\Vc_u}^{\mathrm{max}} = \max_{v \in \Vc_u}\hbar_{v}$
    \For{$h := \hbar_{\Vc_u}^{\mathrm{min}}+1$ to $\hbar_{\Vc_u}^{\mathrm{max}}+1$}
      \State \parbox[t]{8cm}{$\Vc_u^{<h}$ are $u$'s neighbors with hop count less than $h$.}
      \State Select $\Vc_u^\star \subseteq \Vc_u^{<h}$ to maximize $\varrho_{u \rightarrow b}$
             from \eqref{eq:urf_weight2}, \eqref{eq:urf_metric_ref}.
      \State Let $\varrho_{u \rightarrow b}^\star$ be the maximum $\varrho_{u
             \rightarrow b}$.
      \If{$\varrho_{u \rightarrow b}^\star \ge \tau_{k-h+1}$}
        \State $\hbar_u := h$
        \State Add $u$ to $\Vc$. Add links $\{(u,v) : v \in \Vc_u^\star\}$ to $\Ec$.
	\State Break from \textbf{for} loop over $h$.
      \EndIf
    \EndFor %
\EndFor %
\State \textbf{Return}: $G, \vect{\hbar}$
\end{algorithmic}
\end{algorithm}

At each round $k$, a node $u$ decides whether it should join the
routing topology with mesh hop count $\hbar_u$.  If node $u$ joins
with hop count $\hbar_u$, then $u$'s downstream neighbors are the
neighbors $v_i$ with a mesh hop count less than $\hbar_u$ that
maximize $\varrho_{u \rightarrow b}$ from \eqref{eq:urf_metric_ref}.
Node $u$ decides whether to join the topology, and with what mesh hop
count $\hbar_u$, by comparing the maximum reliability $\varrho_{u
  \rightarrow b}^\star$ for each mesh hop count $h \in \{\min_{v
   \in \Vc_u}\hbar_v+1, \ldots, \max_{v \in
    \Vc_u}\hbar_v+1\}$ with a threshold $\tau_m$ that depends on
  $h$.  The threshold $\tau_m$ is selected from a predefined
  vector of thresholds $\vect{\tau} = [\tau_1 \; \cdots \; \tau_M] \in
  [0,1]^M$ using the index $m = k - h + 1$, as shown in
  Figure~\ref{fig:delay_thresh}.  When there are multiple $h$ with
  $\varrho_{u \rightarrow b}^\star \ge \tau_m$, node $u$ sets its mesh
  hop count $\hbar_u$ to the smallest $h$.  If none of the $h$
  have $\varrho_{u \rightarrow b}^\star \ge \tau_m$, then node $u$
  does not join the network in round $k$.

\onefigure{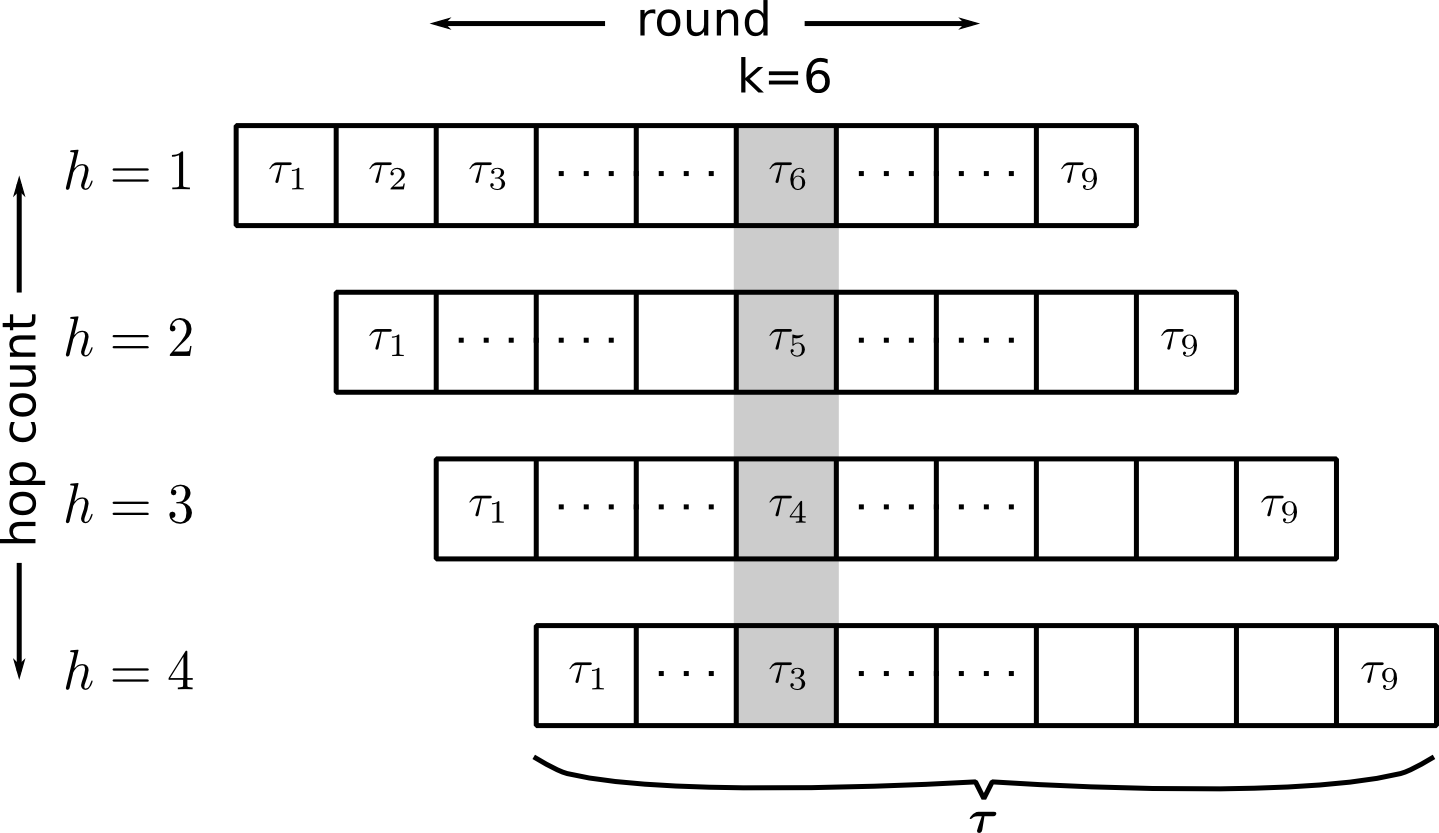}{!}{230pt}
          {Thresholds for mesh hop count}
          {Illustration of how thresholds are used to help assign a
           node a mesh hop count.  The horizontal row of thresholds
           represent $\vect{\tau}$.  The shaded vertical column of
           thresholds are the thresholds tested by a node in round
           $k$.  A node $u$ picks the smallest mesh hop count $h$ such
           that $\varrho_{u \rightarrow b}^\star \ge \tau_m$ (see text
           for details).}
          {fig:delay_thresh}

For the algorithm to work correctly, the thresholds $\vect{\tau}$ must
decrease with increasing $m$.  The network designer gets to choose
$\vect{\tau}$ and the number of rounds $K$ to run the algorithm.
URF-DT can construct a better routing topology if $\vect{\tau}$ has
many thresholds that slowly decrease with $m$, but the algorithm will
take more rounds to construct the topology.

Algorithm~\ref{alg:delay_thresh} is meant to be implemented in
parallel on the nodes in the network.  All the nodes have the vector
of thresholds $\vect{\tau}$.  In each round, each node $u$ listens for
a broadcast of the pair $(\varrho_{v_i \rightarrow b}$, $\hbar_{v_i})$
from each of its neighbors $v_i$ that have joined the routing
topology.  After receiving the broadcasts, node $u$ performs the
computations and comparisons with the thresholds to determine if it
should join the routing topology with some mesh hop count $\hbar_u$.
Once $u$ joins the network, it broadcasts its value of $(\varrho_{u
\rightarrow b},\hbar_u)$.

After a node $u$ joins the network, it may improve its reliability
$\varrho_{u \rightarrow b}$ by adding outgoing links to other nodes
with the same mesh hop count.  To prevent routing loops, a node $u$ may
only add a link to another node $v$ with the same mesh hop count if
$\varrho_{v \rightarrow b} > \varrho_{u \rightarrow b}$, where both
URF metrics are computed using only downstream neighbors with
lower mesh hop count.

\subsection{Discussion}
\label{sec:topo_discuss}

The slowest step in the URF-DT algorithm is selecting the optimal set
of downstream neighbors $\Vc_u^\star$ from the neighbors with hop
count less than $h$ to maximize $\varrho_{u \rightarrow b}$.
Properties~\ref{prop:urf_add_link_lower} and
\ref{prop:urf_add_link_raise} of the URF metric make it difficult to
find a simple rule for selecting downstream neighbors.  Rather than
compute $\varrho_{u \rightarrow b}$ for all possible $\Vc_u^\star$ and
comparing to find the maximum, one can use the following
\emph{lexicographic approximation} to find $\Vc_u^\star$.  First,
associate each outgoing link $(u,v)$ with a pair $(\varrho_{v
  \rightarrow b},p_{uv})$ and sort the pairs in lexicographic order.
Then, make one pass down the list of links, adding a link to
$\Vc_u^\star$ if it improves the value of $\varrho_{u \rightarrow b}$
computed from the links that have been added thus far.  This order of
processing links is motivated by
Property~\ref{prop:increase_urf_downstream} of the URF metric.

Note that the URF metric in the URF-DT algorithm can be replaced by
any metric which can be computed on a node using only information from
a node's downstream neighbors.  For instance, the URF metric can be
replaced by the RRURF metric described in
Section~\ref{sec:urf_discussion}.

\section{Simulations}
\label{sec:simulations}

This section compares the performance of the URF-DT algorithm with two
other simple mesh topology generation schemes described below:
\textsc{Minimum\_Hop} (MinHop) and \textsc{URF-Global\_Greedy}
(URF-GG).  The performance measures are each node's URF metric
$\varrho_{v \rightarrow b}$ and the maximum number of hops from each
node to the sink.

MinHop generates a loop-free minimum hop topology by
building a minimum spanning tree rooted at the sink on the undirected
connectivity graph and then orienting edges from nodes with a higher
minimum hop count to nodes with a lower minimum hop count.  If node
$u$ and $v$ have the same minimum hop count but node $u$ has a smaller
maximum link probability to nodes with a lower hop count, $u$ routes
to $v$.  This last rule ensures that we utilize most of the links in
the network to increase reliability (otherwise, MinHop performs very poorly).

URF-GG is a centralized algorithm that adds nodes sequentially to the
routing topology, starting from the sink.  At each step, every node
$u$ selects the optimal set of downstream neighbors $\Vc_u^\star$ from
nodes that have already joined the routing topology to compute its
maximum reliability $\varrho_{u \rightarrow b}^\star$.  Then, the
node with the best $\varrho_{u \rightarrow b}^\star$ of all nodes that
have not joined the topology is added to $\Vc$, and the links $\{(u,v)
  : v \in \Vc_u^\star\}$ are added to $\Ec$.  Note that URF-GG does
  not generate an optimum topology that maximizes the average URF
  metric across all the nodes (The authors have not found an optimum algorithm.).

Figure~\ref{fig:DAG_batch_comp_URF_MaxHop} compares the performance of
routing topologies generated under the MinHop, URF-DT, and URF-GG
algorithms on randomly generated connectivity graphs.  Forty nodes
were randomly placed in a $10 \times 10$ area with a minimum node
spacing of 0.5 (this gives a better chance of having a connected
graph).  Nodes less than 2 units apart always have a link, nodes more
than 3 units apart never have a link, and nodes with distance between
2 and 3 sometimes have a link.  The link probabilities are drawn
uniformly at random from $[0.7,1]$.  The inputs to URF-DT are the
number of rounds $K = 100$ and a vector of thresholds $\vect{\tau}$
which drops from 1 to 0 in increments of $-0.01$.  We used the
lexicographic approximation to find the optimal set of neighbors
$\Vc_u^\star$.  There were 100 simulation runs of which only 10 are
shown, but a summary of all the runs appears in
Table~\ref{tab:sim_routeTopo_comp}.

{\captionsetup[subfloat]{position=top}
\begin{figure*}
\hspace{-6mm}
\subfloat[]{
  \label{fig:DAG_batch_comp_URF}
  \includegraphics[width=280pt,keepaspectratio=true]
  		  {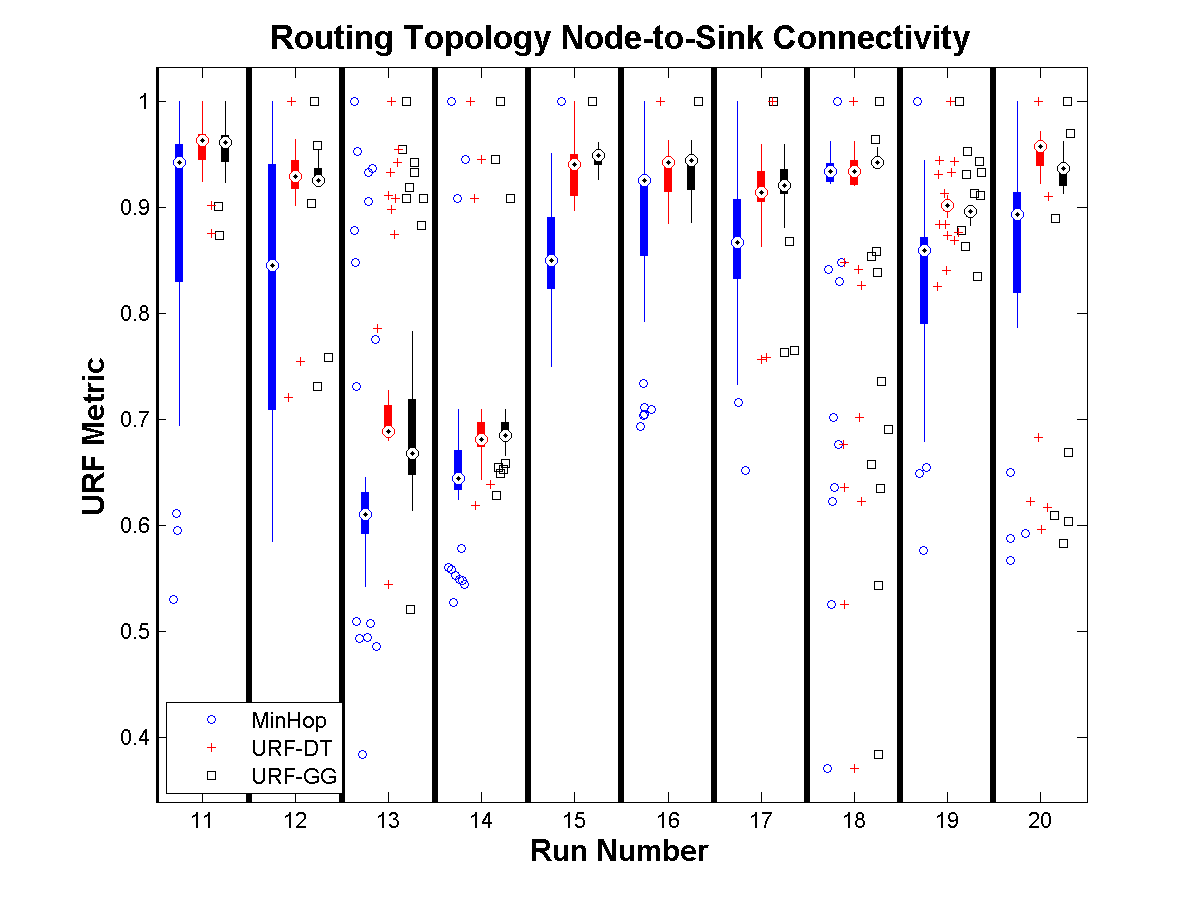}
} \hspace{-10mm}
\subfloat[]{
  \label{fig:DAG_batch_comp_MaxHop}
  \includegraphics[width=280pt,keepaspectratio=true]
  		  {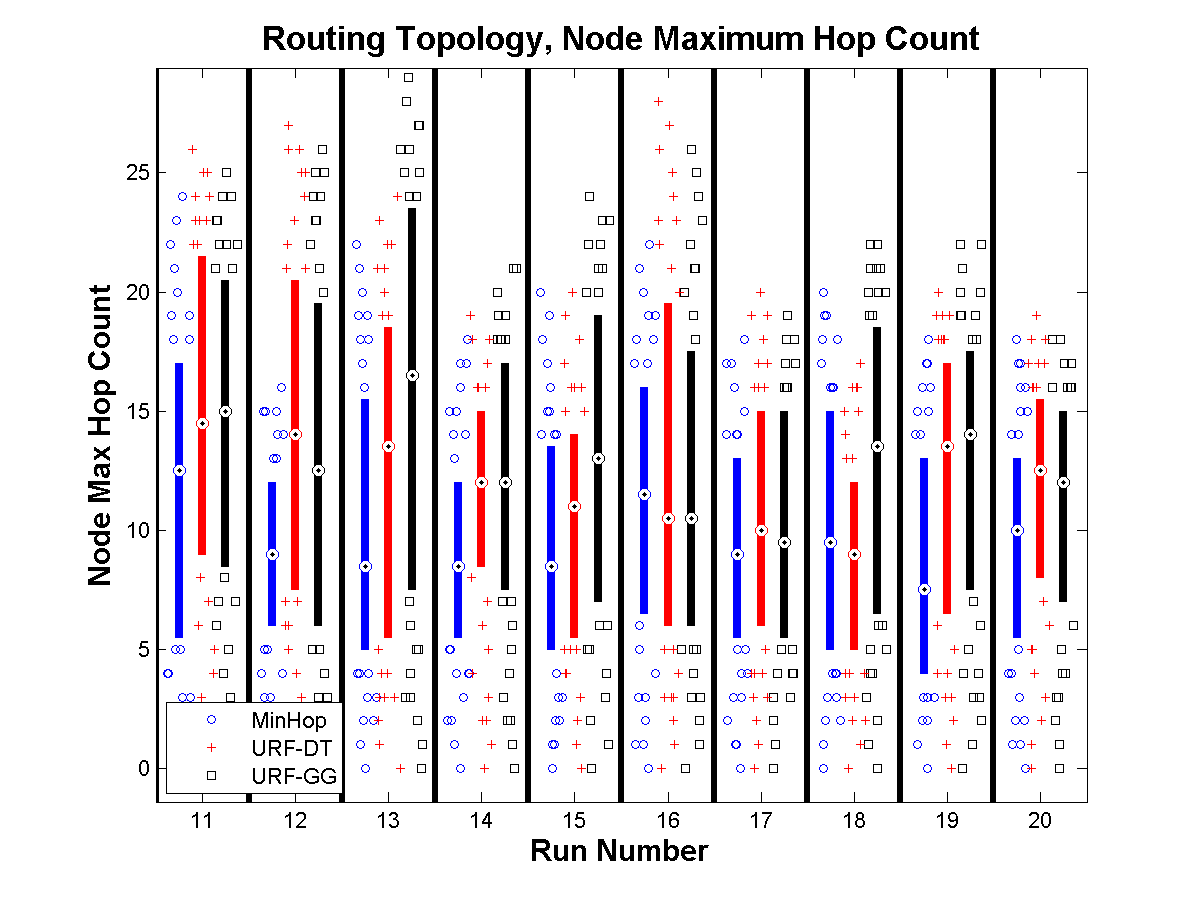}
}
\caption[Comparison of MinHop, URF-DT, and URF-GG DAGs on 10 Random
  Graphs]
{Comparison of routing topologies generated by MinHop, URF-DT, and
  URF-GG, using the \subref{fig:DAG_batch_comp_URF} URF metric
  $\varrho_{v \rightarrow b}$ and \subref{fig:DAG_batch_comp_MaxHop}
  maximum hop count on each node.  The distributions are represented
  by box and whiskers plots, where the median is represented by a
  circled black dot, the outliers are represented by points, and the
  interquartile range (IQR) is 1.5 for \subref{fig:DAG_batch_comp_URF}
  and 0 for \subref{fig:DAG_batch_comp_MaxHop}.}
\label{fig:DAG_batch_comp_URF_MaxHop}
\end{figure*}
}

\begin{table}
\begin{center}
\caption{MinHop, URF-DT, URF-GG Routing Topology Statistics over 100 Random Graphs}
\label{tab:sim_routeTopo_comp}
\begin{tabular}{|r||ccc|cc|}
\hline
Routing  & \multicolumn{3}{c|}{URF Metric $\varrho_{v \rightarrow b}$} &
           \multicolumn{2}{c|}{Max Hop Count}\\
Topology & mean & median & variance & mean & median\\
\hline \hline
  MinHop  & 0.8156 & 0.8252 & 0.0075 & 10.50 & 10.59\\
  URF-DT  & 0.8503 & 0.8539 & 0.0041 & 11.41 & 11.68\\
  URF-GG  & 0.8529 & 0.8549 & 0.0039 & 12.38 & 12.76\\
\hline
\end{tabular}
\end{center}
\end{table}

While in some runs the URF-DT topology shows marginal improvements in
reliability over the MinHop topology, other runs (like run 17) show a
significant improvement.\footnote{A small increase in probabilities
  close to 1 is a significant improvement.}
Figure~\ref{fig:DAG_batch_comp_MaxHop} shows that this often comes at
the cost of increasing the maximum hop count on some of the nodes
(though not always, as shown by run 17).

\section{Conclusions}
\label{sec:conclusions}

Both the FPP and URF metrics show that multiple interleaved paths
typically provide better end-to-end reliability than disjoint paths.
Furthermore, since they were derived directly from link probabilities,
the DAG representing the routing topology, and simple packet
forwarding models, they help us understand when a network is reliable.
Using these routing topology metrics a network designer can estimate
whether a deployed network is reliable enough for his application.  If
not, he may place additional relay nodes to add more links and paths
to the routing topology.  He may also use these metrics to quickly
compare different routing topologies and develop an intuition of which
ad-hoc placement strategies generate good connectivity graphs.

These metrics provide a starting point for designing routing protocols
that try to maintain and optimize a routing topology.  The URF-DT
algorithm describes how to build a reliable static routing topology,
but it would be interesting to study algorithms that gradually adjusts
the routing topology over time as the link estimates change.

\bibliographystyle{IEEEtran}
\bibliography{techrep11-routeTopo,IEEEtranBSTcontrol}

\appendix
\vspace{-5mm}
\begin{align*}
      \varpi_{uv} &= \sum_{\Ec' \in 2^{\Ec_u \backslash l}} 
        \frac{p_l}{|\Ec'|+1} \left( \prod_{e \in \Ec'} p_e \right)
        \left( \prod_{\bar{e} \in \Ec_u \backslash (\Ec' \cup l)}
	       1 - p_{\bar{e}} \right)\\
        &= p_l \sum_{\Ec' \in 2^{\Ec_u \backslash l}} 
           \frac{1}{|\Ec'|+1} \left( \prod_{e \in \Ec'} p_e \right)
           \left( \prod_{\bar{e} \in \Ec_u \backslash (\Ec' \cup l)}
	          1 - p_{\bar{e}} \right)\\
	&= p_l \int_0^1 \sum_{\Ec' \in 2^{\Ec_u \backslash l}} 
           \left( \prod_{e \in \Ec'} p_e \right)
           \left( \prod_{\bar{e} \in \Ec_u \backslash (\Ec' \cup l)}
	   1 - p_{\bar{e}} \right)x^{|\Ec'|}dx\\
	&= p_l \int_0^1 \prod_{e \in \Ec_u \backslash l} 
           \left( (1 - p_e) + p_e x \right)dx\\
	&= p_l \int_0^1 \prod_{e \in \Ec_u \backslash l} 
           \left( 1 - p_e (1 - x) \right)dx\\
	&= p_l \int_0^1 \prod_{e \in \Ec_u \backslash l} 
           \left( 1 - p_e x \right)dx\\
\end{align*}

\end{document}